\documentclass[12pt]{nature}

\usepackage{graphicx, amsmath, color, url}

\usepackage{times}

\title{Mapping of shadows cast on a protoplanetary disk
by a close binary system} 

\author
{V. D'Orazi, R. Gratton, S. Desidera, H. Avenhaus, D. Mesa, T. Stolker, 
E. Giro, S. Benatti, H. Jang-Condell, E. Rigliaco, E. Sissa, T. Scatolin, M. Benisty, T. Bhowmik, 
A. Boccaletti, M. Bonnefoy, W. Brandner, E. Buenzli, G. Chauvin, S. Daemgen, M. Damasso, M. Feldt, 
R. Galicher, J. Girard, M. Janson, J. Hagelberg, D. Mouillet, Q. Kral, J. Lannier, A. M. Lagrange,
M. Langlois, A.-L. Maire, F. Menard, O. Moeller-Nilsson, C. Perrot, S. Peretti, P. Rabou, J. Ramos, L. Rodet, R. Roelfsema, 
A. Roux, G. Salter, J. E. Schlieder, T. Schmidt, J. Szulagyi, C. Thalmann, P. Thebault, G. van der Plas, 
A. Vigan, A. Zurlo}
\begin{document} 
\maketitle

\clearpage

{\bf For a comprehensive understanding of planetary formation and evolution, we need to investigate the environment
in which planets form: circumstellar disks. Here we present
high-contrast imaging observations of V4046 Sagittarii, a
20-Myr-old close binary known to host a circumbinary disk.
We have discovered the presence of rotating shadows in the
disk, caused by mutual occultations of the central binary.
Shadow-like features are often observed in disks\cite{garufi,marino15}, but those
found thus far have not been due to eclipsing phenomena.
We have used the phase difference due to light travel time to
measure the flaring of the disk and the geometrical distance
of the system. We calculate a distance that is in very good
agreement with the value obtained from the Gaia mission's
Data Release 2 (DR2), and flaring angles of $\alpha = 6.2 \pm 0.6 $ deg
and $\alpha = 8.5 \pm 1.0 $ deg for the inner and outer disk rings, respectively. Our technique opens up a path to explore other binary
systems, providing an independent estimate of distance and
the flaring angle, a crucial parameter for disk modelling.}

The stellar system under scrutiny here is peculiar. V4046 Sgr
(HD 319139) is a double-lined spectroscopic binary with an
orbital period of P=2.42 days\cite{stempels04}.
The K-type stars that comprise the binary system have nearly equal masses, M$_{\star,A}$=0.90$\pm$0.05 M$_\odot$ and M$_{\star,B}$=0.85$\pm$0.04 M$_\odot$, a separation of $a$=0.045 au, and an eccentricity e$<$ 0.001 (i.e., a circular orbit\cite{rosenfeld13}). V4046 Sgr is a proposed member of the $\beta$ Pic Moving Group \cite{torres08} with a kinematic distance of $d$=73 pc \cite{torres06}. The estimated dynamical masses of the components along with their temperatures and luminosities suggest an age of $\approx$ 10 $-$ 20 Myr.  
At this relatively advanced age, gas-rich disks have usually already been dissipated\cite{wyatt08}.
Several mechanisms have been advocated in order to explain the survival of the circumbinary disk orbiting V4046 Sgr, such as e.g., tidal torque induced by the close binary on its disk 
that inhibits accretion and increases lifetime with respect to a single star\cite{alexander12}. Another possibility is the presence of a cavity that may limit accretion flows onto the central star\cite{rosenfeld13}. This cavity could be the result of dynamical interactions with an unseen massive companion with a mass limit of M$<$0.07 M$_\odot$ at separations $a > 2.9~au$ (see Figure 8 in \cite{rosenfeld13}). Alternatively, 
the gap could instead be related to photoevaporating winds that create large pressure traps (see \cite{rosenfeld13} for further details). 

Given the intriguing nature of this object, multiple multi-wavelength observations have been carried out. This  includes H$_{\alpha}$ emission \cite{merrill50} and sub-millimeter studies \cite{jensen96}.
 Rosenfeld and collaborators \cite{rosenfeld13} reported a large inner hole ($r=29$ au) that is spatially resolved in 1.3 mm continuum emission. 
 In their work they estimated a dust+gas mass for the disk of 0.094 M$_{\odot}$ and derived a disk inclination of $i$=33.5$^{+0.7}_{-0.4}$, with a position angle (P.A.) of 76 degrees.
Subsequently, GPI [Gemini Planet Imager\cite{macintosh14}] observations revealed a relatively narrow ring of polarised NIR flux whose 
brightness peaks at $\sim$14 au\cite{rapson15b}. This $\sim$14 au radius ring is surrounded by a fainter outer halo of scattered light extending to roughly 45 au, 
which coincides with the previously detected millimeter-wave thermal dust emission.

In this study we have used the ESO-VLT facility SPHERE\cite{beuzit} instrument and acquired IFS\cite{claudi} and IRDIS\cite{dohlen} high-contrast imaging and spectroscopic near-infrared observations (see Methods). 
Figure 1 displays the wavelength collapsed IFS images for the two observed epochs in 2015 and 2017 after a PCA (Principal Component Analysis) algorithm was applied to remove the quasi-static speckle contribution. Observations show a double-ring structure, similar to that reported by \cite{rapson15b} from GPI datasets. Moreover, we
confirm the presence of shadow-like features that were previously suggested \cite{rapson15b}.  We found that the shadows have rotated by 11$\pm$1 degrees between the observed
epochs, in agreement with what is expected from the central binary phase (see Table 1). 
We are observing the ``{\it penumbra}'' created when the primary component of the binary system 
partially eclipses the secondary star and causes a reduction of the stellar flux illuminating the disk surface (see Figure 2).  We note that the total eclipse of less than 0.5 deg would be too narrow to be detected. It is noteworthy that the presence of shadows is expected because the binary orbital plane and the disk plane are aligned, with a difference in inclination of only $dI=0\pm$1 degree \cite{rosenfeld12}.
Thanks to the high-contrast imaging capabilities of SPHERE we are able to detect these features that were previously not observable because of technical limitations.

High-resolution observations of protoplanetary disks have
revealed the presence of small-scale features including spirals,
cavities and shadows\cite{benisty17}. Shadows are not unusual on the surfaces
of circumstellar disks, but the physical mechanisms responsible for previously detected features differ from those at work here.
Examples of alternative astrophysical mechanisms include a self-shadowing
process\cite{garufi} due to different inclinations of the inner and
outer rings, and the presence of circumstellar material in binary systems
(the case of GG Tauri\cite{itoh14}). Other well-known examples include
HD 142527\cite{marino15}, MP Musca\cite{wolff16} and TW Hydrae\cite{debes17}. We also refer to
refs 21$-$23 and references therein for extensive diskussions on this
topic. In contrast, our observations provide evidence of fast-moving
shadows in a circumbinary disk cast by the central binary system.
Crucially, we can exploit the presence of these shadows to infer two fundamental properties of the system: the flaring angle
of the disk and the distance to the system. The distance determination
is a purely geometrical estimate that is completely independent
from other methods. We have calculated the expected phase
delay of the shadow, due to the light travel time, with respect to the
binary phase, and compared the theoretical (expected) value with
our observed measurements. This has been done for the inner disk (at 13 au) and the outer disk (at 29 au) and for shadows on the near side and on the far side of the disk.
The phase delay of the shadow can be calculated as explained in Methods; here we recall that 
phase delay depends on the binary phase ( $\phi$), the flaring and inclination angles of the disk ($\alpha$ and $i$, in radians), the distance between the stars and the disk rings ($r$, where the inner ring is at 13 au and outer ring at 29 au), the speed of light $c$, and the central binary period ($P$) in seconds. 
 In our Equation 1, there is a dependence of the phase delay on the system distance (which enters into the ring positions as expressed in au) and the disk flaring angle. 

To break the flaring-distance degeneracy, we have used polarimetric differential imaging (PDI) observations acquired with
IRDIS in 2016\cite{avenhaus18} to estimate the disk flaring.  Observations exploiting differential imaging techniques (e.g., angular differential imaging,  PCA) help in identifying the shadow feature, because they naturally enhance discontinuities in the images. On the other hand, they do not reproduce the correct (real) depth of the shadow given the self-subtraction effects produced by the image post-processing technique. After performing rotation, stretching, and transformation to polar coordinates of the reduced PDI H band image, 
we performed a local polynomial fit to the ring brightness profile and measured the depth of the shadows. The corresponding uncertainty on this measurement is given 
by the r.m.s. of the fit corrected for small-number statistics.  
Note that we have assumed that the polarisation does not change across the shadow and the drop in polarised intensity is solely due to a drop in intensity.
The positions of the shadows as observed in the polarimetric images can be easily identified in Figure~3. The described procedure was performed for both the inner and outer rings while 
averaging the shadow depths for the near side and far side of the disk. Specifically, for the inner disk we estimated depths of 0.14$\pm$0.02 and 0.17$\pm$0.03 
for the near side and the far side, respectively. In the outer ring we measure 0.16$\pm$0.06 and 0.10$\pm$0.05 (errors are r.m.s of the fit corrected for small-number statistics).  The mean values result in 0.15$\pm$0.02 and 0.13$\pm$0.04, respectively.

To translate the observed shadow depth into disk flaring measurements, we calculated the illumination of the disk from the central binary as a function of the angle with respect to the shadow centre. Our computation assumes that the illumination depends on the sum of the areas of the two stars as seen from the rings (the effects of limb darkening were neglected), and that the two stars have the same effective temperatures. Evaluations of the impact of this assumption on the final result were found to be negligible. We then calculated the shadow depth for $\alpha$ between 0 and 15 degrees with steps of 1 degree (see Supplementary Figure 2). As expected, 
 the depth of the shadow is directly related to the flaring of the disk because the flatter the disk (i.e., less flared), the deeper the shadow. 
 Note that for values $>$14 degrees, no shadows would have been detectable in our dataset, because we only see the disk surface in scattered light.
From comparison with our measurements on the PDI data, the interpolation of our relationship shown in Supplementary Figure 2 provides a flaring angle of 
 $\alpha$= 6.2$\pm$0.6 deg and $\alpha$=8.5$\pm$1.0 deg for the inner and outer rings, respectively. We include in this estimate the impact of the instrumental spatial resolution (FWHM=50 mas).
 Our values can be compared with estimates by \cite{avenhaus18}, who have determined
$\alpha$=5.33$\pm$0.34 deg for the inner disk and 7.45$\pm$0.23 deg for the outer ring. In their study, Avenhaus and collaborators performed a fit of the two bright rings, assuming that 
they represent the disk scattering surface and that the circular rings are inclined and displaced from the mid-plane. The errors were estimated using an MCMC algorithm. 
Despite the differences in the procedures, there is very good agreement between the two flaring angle determinations. The flaring angle determined for V4046 Sgr is generally in agreement with that measured by \cite{avenhaus18} for other classical T Tauri stars and the marginally lower value is likely to be because of the more advanced evolution of this disk.

Including our flaring estimates in Equation 1, we can now determine which distance values minimise the difference between the expected and measured positions of the shadows.
We found that the difference between the expected shadow position, calculated by applying the delay due to light travel time to the binary phase, and that measured on our data,
(see Table 1) is  2.86$\pm$0.18 deg where the error is the r.m.s from the average of the shadows observed in the disk near side (because if their higher accuracy).
This corresponds to a diskrepancy in the binary phase of 7$\times 10^{-3}$. However ,our error in the phase is much larger, i.e., 0.03.
Because of this large uncertainty, we decide to exploit the phase difference between inner and outer rings, which is independent of the binary phase, to infer the system distance. 
In this way we obtain a pure geometrical distance of d=78$\pm$8 pc, which is in good agreement with the new Gaia DR2 distance of d=72.4$\pm$0.3 pc.

Finally, we emphasise that the presence of shadows in the PDI observations (Figure~3) demonstrates that these features are real and not an artefact due to post-processing techniques such as e.g., ADI and similar algorithms that can introduce artificial structures.
To quantify this, we carried out an ADI simulation to test the possible effects of these procedures and see if they can introduce spurious shadow-like structures. Our simulation results suggest that artefacts introduced by ADI post-processing are significantly different 
from the shadows detected in our observations and they are preferentially located along the minor axis of the disk (see Methods).

Our study furnishes two main results. Firstly, we provide an estimate of the flaring 
angle by using a new approach that does not depend on the details of disk modelling.
Adopting the formalism given in \cite{rosenfeld13}, we found values of the flaring angles that correspond to $\psi$=1.5 (in their Equation 2): this is larger than the standard value of $\psi$=1.25 usually assumed for circumstellar disks (see also \cite{rosenfeld13} for a similar result). Consequently, our finding provides quite stringent observational constraints on disk modelling.
Moreover, we obtain a measurement of the system distance that is complementary and independent from other methods. On the other hand, very accurate distances available from Gaia can be used to determine the disk flaring on much more solid grounds, because only one free parameter is then considered in our calculations.
Our techniques can be applied to other systems, once the conditions of the close binary system and alignment of the binary and disk planes are understood. In this regard, future observations of 
systems like $DK$ Tau, $AK$ Sco, and new SB2 stars discovered in Sco-Cen\cite{kuruwita18} will be crucial to study further.

\section*{Correspondending author}
Correspondence to Valentina D'Orazi

\section*{Acknowledgments}
The authors thank the ESO Paranal Staff for support for conducting the observations.  E.S., R.G., D.M., S.D. acknowledge support from the "Progetti Premiali" funding scheme of the Italian Ministry of Education, University, and Research. D.M. acknowledges support from the ESO-Government of Chile Joint Committee
program ``Direct imaging and characterization of exoplanets''. E.R. is supported by the European Union's Horizon 2020 research and innovation programme under the Marie Sk\l odowska-Curie grant agreement No 664931. This work has been supported by the project PRIN-INAF 2016 The Craddle of Life - GENESIS-SKA (General Conditions in Early Planetary Systems for the rise of life with SKA). A.Z. acknowledges support from the CONICYT + PAI/ Convocatoria nacional subvenci\'on a la instalaci\'on en la academia, convocatoria 2017 + Folio PAI77170087. The authors acknowledge financial support from the Programme National de Plan\'{e}tologie (PNP) and the Programme National de Physique  Stellaire (PNPS) of CNRS-INSU. This work has also been supported by a grant from the French Labex OSUG@2020 (Investissements d'avenir - ANR10 LABX56). The project is supported by CNRS, by the Agence Nationale de la Recherche (ANR-14-CE33-0018). This work is partly based on data products produced at the SPHERE Data Centre hosted at OSUG/IPAG, Grenoble. We thank P. Delorme and E. Lagadec (SPHERE Data Centre) for their efficient help during the data reduction process. SPHERE is an instrument designed and built by a consortium consisting of IPAG (Grenoble, France), MPIA  (Heidelberg, Germany), LAM (Marseille, France), LESIA (Paris, France), Laboratoire  Lagrange (Nice, France), INAF Osservatorio Astronomico di Padova (Italy),  Observatoire de Gen\`{e}ve (Switzerland), ETH Zurich (Switzerland), NOVA (Netherlands), ONERA (France) and ASTRON (Netherlands) in collaboration with ESO. SPHERE was funded by ESO, with additional contributions from CNRS (France), MPIA (Germany), INAF (Italy), FINES (Switzerland) and NOVA (Netherlands). SPHERE also received funding from the European Commission Sixth and Seventh Framework Programmes as part of the Optical Infrared Coordination Network for Astronomy (OPTICON) under grant number RII3-Ct-2004-001566 for FP6 (2004-2008), grant number 226604 for FP7 (2009-2012) and grant number 312430 for FP7 (2013-2016). 
This work made extensive use of the SIMBAD and NASA ADS databases.

\noindent
{\bf Author contributions}

\noindent
V. D'Orazi, R. Gratton, S. Desidera, H. Avenhaus, T. Stolker, E. Giro, E. Rigliaco, E. Sissa, T. Scatolin analysed the SPHERE data and prepared the manuscript and the figures.
D. Mouillet, W. Brandner, D. Mesa, G. Salter, A. Boccaletti, and A.M. Lagrange performed the observations in 2015 and 2017. 
T. Bhowmik, E. Buenzli, S. Daemgen, R. Galicher, J. Hagelberg, J. Lannier, M. Langlois, A.-L. Maire, S. Peretti, C. Perrot, L. Rodet, and T. Schmidt 
reduced the data as part of the SPHERE Data Reduction team in the framework of the SPHERE GTO.
S. Benatti and M. Damasso performed the orbital period analysis.
M. Benisty, A. Boccaletti, M. Bonnefoy,  J. Girard, M. Janson, Q. Kral, F. Menard. J. E. Schlieder,  J. Szulagyi, C. Thalmann, P. Thebault, G. van der Plas, A. Vigan, A. Zurlo
contributed to the diskussion of the results with helpful and fundamental comments and suggestions.
G. Chauvin, M. Feldt are key people of the SHINE survey carried out with SPHERE. O. Moeller-Nilsson, R. Roelfsema, P. Rabou, A. Roux,  J. Ramos are SPHERE builders

\noindent
{\bf Author information}

{\bf Affiliations}

\noindent
{\it INAF Osservatorio Astronomico di Padova, vicolo dell'Osservatorio 5, 35122, Padova, Italy} \\
V. D'Orazi, R. Gratton, S. Desidera, D. Mesa, E. Giro, S. Benatti, E. Rigliaco, E. Sissa, T. Scatolin. \\
{\it Max Planck Institute for Astronomy, Koenigstuhl 17, 69117, Heidelberg, Germany}\\
H. Avenhaus, W. Brandner, M. Feldt, A.-L., Maire, O. Moeller-Nilsson, J. Ramos, J. E. Schlieder\\
{\it Institute for Particle Physics and Astrophysics, ETH Zurich, Wolfgang-Pauli- Strasse 27, 8093, Zurich, Switzerland}\\
H. Avenhaus, T. Stolker, E. Buenzli, S. Daemgen, C. Thalmann\\
{\it Department of Physics \& Astronomy, University of Wyoming, Laramie, WY 82071, USA.}\\
H. Jang-Condell\\
{\it INCT, Universidad De Atacama, calle Copayapu 485, Copiap\'o, Atacama, Chile}\\
D. Mesa\\
{\it  Universit\'e Grenoble Alpes,CNRS, IPAG, 38000 Grenoble, France}\\
M. Benisty, M. Bonnefoy, G. Chauvin, J. Hagelberg, D. Mouillet, J. Lannier, A.-M. Lagrange, F. Menard, P. Rabou, L. Rodet, A. Roux
G. van der Plas.\\
{\it Instituto de Fisica y Astronomia, Facultad de Ciencias, Universidad de Valparaiso, Av. Gran Bretana 1111, Valparaiso, Chile}\\
C. Perrot\\
{\it LESIA, Observatoire de Paris, Sorbonne Universit\'{e}, Univ. Paris Diderot, Sorbonne Paris Cit\'{e}, 5 place Jules Janssen, 92195 Meudon, France}\\
T. Bhowmik, A. Boccaletti, R. Galicher, J. Girard, Q. Kral, P. Thebault\\
{\it Department of Astronomy, Stockholm University, 114 19 Stockholm, Sweden}\\
M. Janson\\
{\it Institute of Astronomy, University of Cambridge, Madingley Road, Cambridge CB3 0HA, UK}\\
Q. Kral\\
{\it CRAL, UMR 5574, CNRS, Universit\'e Lyon 1, Avenue Charles Andr\'e, 69561 Saint Genis Laval Cedex, France}\\
M. Langlois\\
{\it  Aix Marseille Université, CNRS, Laboratoire d'Astrophysique de Marseille, UMR 7326, 13388, Marseille, France}\\
M. Langlois, G. Salter, T. Schmidt, A. Vigan, A. Zurlo.\\
{\it Observatoire Astronomique de l'Universit\'e de Gen\`eve, 51 Ch. des Maillettes, 1290 Versoix, Switzerland}\\
S. Peretti\\
{\it NOVA Optical Infrared Instrumentation Group, Oude Hoogeveensedijk 4, 7991 PD Dwingeloo, The Netherlands}\\
R. Roelfsema\\
{\it Exoplanets and Stellar Astrophysics Laboratory, NASA Goddard Space Flight Center, Greenbelt, MD 20771, USA}\\
J. E. Schlieder\\
{\it Center for Theoretical Astrophysics and Cosmology, Institute for Computational Science, University of Z\"urich Winterthurestrasse 190, CH-8057 Z\"urich, Switzerland}\\
J. Szulagyi\\
{\it INAF Osservatorio Astronomico di Brera, via Emilio Bianchi 46, 23807 Merate (LC), Italy}\\
E. Giro\\
{\it Space Telescope Science Institute, 3700 St Martin Drive, Baltimore, MD, USA}\\
J. Girard\\
{\it Dipartimento di Fisica e Astronomia, Universit\'a di Padova, via Marzolo 8, 35121, Padova, Italy.}\\
T. Scatolin\\
{\it INAF  Osservatorio Astrofisico di Torino, Via Osservatorio 20, 10025, Pino Torinese (TO), Italy}\\
M. Damasso \\
{\it N\'ucleo de Astronom\'ia, Facultad de Ingenier\'ia y Ciencias, Universidad Diego Portales, Av. Ejercito 441, Santiago, Chile}\\
A. Zurlo\\
{\it Escuela de Ingenier\'ia Industrial, Facultad de Ingenier\'ia y Ciencias, Universidad Diego Portales, Av. Ejercito 441, Santiago, Chile}\\
A. Zurlo \\

\noindent
{\bf Competing Interests}
\noindent
The authors declare no competing interests.

\begin{methods}

 \subsection{SPHERE observations}\label{sec:sphere}

 The current dataset for IFS+IRDIS observations was acquired in the framework of guaranteed time observations (GTO) for the SHINE survey [SpHere INfrared survey for Exoplanets\cite{chauvin17}] with SPHERE at the VLT. 
 The configuration allows for simultaneous acquisition of a low resolution IFS (R$\sim$35) spectrum in the YJ band, covering 0.96-1.34 $\mu$m, or the YH band, covering 0.97-1.66 $\mu$m, while IRDIS is operating in dual-band imaging mode using either two narrow H or two narrow K bands \cite{vigan10}.
 For the first epoch (May 2015), IFS was operated in $YJ$  mode and IRDIS with the $H_2H_3$ (central wavelengths $H_2$=1.59$\mu$m, $H_3$=1.67$\mu$m) dual band filters. For the 2017 observations the $YH$ (IFS) and 
 $K_1K_2$ (IRDIS, $K_1$=2.10$\mu$m, $K_2$=2.25$\mu$m) extended set-up was adopted.  The resulting data-cubes were comprised of 60 and 64 science frames, for 2015 and 2017 respectively, with 64s of integration time. 
 We used the SPHERE Data Reduction and Handling (DRH) pipeline \cite{pavlov} to perform background subtraction, flat field correction, bad pixel removal, 
 and to center the star behind the coronagraph. The DRH pipeline was also use for IFS spectral extraction. The data were reduced by the SPHERE Data Centre \cite{delorme17} and astrometrically calibrated following the methods in \cite{maire16}.
 Along with the science frames, the sequence also includes: $(i)$ Several observing frames that contain satellite sports distributed symmetrically around the central star. The satellite spots are created by applying a waveform to the deformable mirror to create four echoes of the source point spread function. They allow for an accurate determination of the location of the star behind the coronagraph, and (ii) Exposure frames obtained by moving the star's position out of the coronagraph to acquire unsaturated stellar images to perform accurate flux calibration. 
 
 The reduced datacubes were then processed through several Angular Differential Imaging (ADI\cite{marois06}) techniques, including the TLOCI-ADI algorithm (as described in \cite{galicher11}) 
 and a PCA method (which combines ADI and spectral differential imaging) following the prescriptions in \cite{mesa15}.  In Supplementary Figure 1 the outcome of the TLOCI-ADI \cite{galicher18} post-processing for the K-band IRDIS observations is displayed. In Table 1 we report for each observing dataset the barycentric julian date, the corresponding binary phase, and the shadow positions as measured on the inner and outer ring for the near side of the disk in the IFS images.
 
  Polarimetric Differential Imaging (PDI) data were collected as part of the ESO Program 096.C-0523(A) (PI H. Avenhaus) using IRDIS in both broad-band J and H filters.
Observations were carried out during six nights from March 10 2016, to March 15 2016. V4046 Sgr is one of the targets comprising the sample of T Tauri stars with disks presented in 
Avenhaus et al. \cite{avenhaus18}. We refer to this work for the full description of the observing sequence (flux, centering and science frames), exposure times and data reduction, which was performed as in \cite{avenhaus14}.

\subsection{Binary orbit determination}
We have used the INTEGRAL Optical Monitoring Camera (OMC) archive (\url{http://sdc.cab.inta-csic.es/omc/index.jsp}) and calculated the orbital period by using 150 photometric 
measurements that span a time range between March 2003 and October 2012. By adopting the zero point of \cite{quast00}, we provide a time coverage of 30 years.
Our ephemerides result in:

$$
JD=2 ~446 ~998.335 + 2.42129516 \times Y
$$

where $Y$ is the fractional part of the orbital phase $\phi$.
The formal error on our period determination, as given from the sinusoidal fit to our dataset, is 3.122$\times 10^{-5}$ days.

Our estimate is in very good agreement with the period ($2.421296\pm0.000001$ days) derived in ref. \cite{donati11}.
As already noticed in ref.\cite{donati11} this value is several sigma out of the determination by \cite{quast00} and \cite{stempels04};
this might be explained as due to the stellar activity that produces an uncertainty much larger than the nominal error in periodograms.

Given the uncertainty in the period, and the time elapsed since the last observations by \cite{donati11} and our dataset, we conclude that the 
binary phase at the epoch of our observations has an uncertainty of $\pm$0.03.

\hspace{2cm}

 \subsection{Results of the ADI simulation}\label{sec:simulation}

The ADI technique allows for the enhancement of weak features that are responsible for 
azimuthal gradients in high contrast images, making them detectable above the static (instrumental) speckle noise. 
The shadows considered in this paper match this criterion. However, ADI may also create false alarms due to random 
distributions of the noise in the residuals\cite{milli12}. In addition, ADI may cancel real
features out, which have weak or non-azimuthal gradients, such as the emission around the main axis of the observed ellipses that result from the line of sight projection of circular rings. We note here that ADI attenuation is inversely 
proportional to the second derivative of the signal along the azimuth, and hence the signal attenuation due to ADI is expected to be largest
close to the minor axis. While the shadows we found on the disk of V4046 Sgr do not align with either the major or minor
axis of the disk projection, one may wonder whether their prominent appearance in ADI images might be a consequence of the
properties of image post-processing.

To test this issue, we built a forward model simulating the photometric properties of the V4046 Sgr disk and added
a realistic noise model to it  using real data of another target lacking a prominent disk and observed
in conditions similar to those of V4046 Sgr. We then processed the resulting simulated imaging data set through ADI with exactly the same procedure and parameters used on our observed data of V4046 Sgr.

To estimate a photometric model for the disk, we computed a geometric model under the hypothesis
that it can be represented by a solid scattering surface. Similar approaches are described in \cite{stolker16}, \cite{maire17}, and \cite{sissa18} for the case of HD 100546 and SAO 206462, respectively. The surface is assumed to have cylindrical symmetry and its height above the mid-plane is described by the law: $h=c1*(r/r_0)^b$, where c1 is a constant and r the distance to the star. We further assumed that scattering efficiency is represented by a 
two-component Henyey-Greenstein (HG) function \cite{henyey} with coefficients as defined in \cite{milli17} for the case of HR 4796. The model includes detailed consideration of both disk walls and disk self-shadowing.
In practice, we used the following model parameters: a distance of 78 pc (this paper); disk inclination and position angle 
of 33.5 and 76 degrees, respectively \cite{rapson15b}; two disks, the first one between 11 and 18 au and the second one
between 21 and 38 au; and thickness parameters $c1=0.042$, $b=1.36$, and $r_0=1$ au. These values are compatible with the
results for the flaring angles we obtained with our modelling of the shadows. A Gaussian smoothing is then applied 
to reproduce the instrumental resolution, and the image is then multiplied by the coronagraph transmission (A. Boccaletti, private communication).  
Final images are produced with scale and size corresponding to IFS and IRDIS images so that they 
can be directly summed to the datacube of a star without a prominent disk. For this we used an observation of HD95086 
obtained with SPHERE on April 2017. HD95086 has similar magnitude to V4046 Sgr.  The total exposure time and the field 
rotation of the images are also similar. Several relatively small adjustments of the disk intensity were needed to take into account
the residual differences. 

In Supplementary Figure 3 we compare the result of this ADI simulation (left-hand panel) with the real ADI reduction of V4046 Sgr (right-hand panel). The figure reveals that ADI actually creates gaps in the simulated disk that are not present in the original model. However, the most
prominent gap in the simulation is located along the minor axis and it is much wider than that expected for a shadow. Thus, the shadow 
properties are significantly different from the gap observed in the ADI image.

\hspace{2cm}

 \subsection{Calculations of the phase delay}\label{sec:pd}

In order to compute the phase delay between the shadow and the binary due to the light travel time we first considered a geometric sketch of the system as displayed in Supplementary Figure 4. The parameters that have to be included to calculate the phase delay are the inclination of the disk (i=33.5 deg), the distance of the system, which we take into account when converting the positions of the rings from angular to scalar distances, and the flaring angle $\alpha$. 
The locus of points on the surface of an axisymmetric disk with a 
flaring angle $\alpha$ at a distance $r$ from the central star is a ring of radius $rcos\alpha$ at a distance $rsin\alpha$ above the
star. We can orient our coordinate axis so that the star is at the origin and the observer
is toward the positive z-axis. If the disk is face-on with respect to the observer, then
the equation for this ring can be expressed parametrically as
\vspace*{-0.5cm}
\begin{equation*}
x'=r~cos\alpha~cos\phi ; ~~~~~~ y'=r~cos\alpha~sin\phi ; ~~~~~~~ z'=r~sin\alpha,
\end{equation*}
where $\phi$ is the phase angle representing the location along the ring. $\phi$ is defined clockwise starting from the East side of the major axis of the apparent disk ellipse (see Figure 1).

In order to account for the inclination, we performed a coordinate transformation representing a rotation about the $x$ axis.

The difference in the light path is then $dx = r - z$, because the light has to travel first to the point on the ring and then scatter along the z-axis to the observer.
Thus,
\vspace*{-0.4cm}
\begin{equation*}
dx=r[1-~\sin{\phi}\cos{\alpha}\sin{i} - \sin{\alpha}\cos{i}].
\end{equation*}

When sin$\phi$=1 (near side of the disk), this produces  $dx=r - r \sin{(i+\alpha)}$, whereas for sin$\phi=-1$ (far side) we have $ dx=r+r\sin{(i-\alpha)}$.
We finally obtain that the phase difference (PD) is:
\vspace*{-0.2cm}
\begin{equation}
PD=\frac{r}{c  P} ~360^o [1-\sin{\alpha}  ~\cos{i}  - \sin{\phi} \sin{i}  \cos{\alpha}]. 
\end{equation}
Thus, the distance d (in parsec) is:
$$
d = \frac{PD/360}  {f(\alpha, \phi, i) \times s} \times c \times P \times \cos {\alpha} 
$$

where s is the semi-major axis of the disk in arcseconds and $f=[1-\sin{\alpha}  ~\cos{i}  - \sin{\phi} \sin{i}  \cos{\alpha}] $.

We note that the difference in the position angles obtained for the inner and outer disks (see Figure 3) due to the light travel time delay indicates that the binary is rotating counter-clockwise. This agrees with the phase difference for the binary between the 2015 and 2017 observing epochs.
We have compared this Equation with those reported in \cite{kama16}, where the authors consider the effect of light travel time on the shape of a shadow cast by a clump orbiting
close to central star.
In our calculation we have assumed that the separation of the two components is negligible 
with respect to the disk scale. This implies that their Equation 5 is equal to zero.
If we consider a different definition for the phase, their Equation 10 is then exactly the same as our Equation 1.

\hspace{2cm}

\subsection{A sanity check on the shadow depth calculations}\label{sec:check}

As mentioned in the main body of the manuscript, the shadow depth (we recall that we detect the penumbra due to central binary eclipse) 
has been determined by exploiting the PDI data. 

However, simple calculations can give further corroboration to our results. The full width half maximum (FWHM) of the penumbra is approximately 20 degrees. 
If we assume the same stellar radius for both components (R=1.2 R$_{\odot}$), basic geometrical calculations provide that they should produce a penumbra of $\approx$26 degrees, 
given $\alpha$ = FWHM/2, and $\tan(\alpha)=\frac{R_*}{\frac{a}{2}}$, where R$_*$ is the stellar radius and $a$ is the separation. 
The value of 26 degrees is in agreement with a flaring angle of $\sim$ 5 deg, which is very close to our determination of flaring angle given by the shadow depth measurements in the SPHERE images.

Thus, this basic cross check provides supporting arguments for our estimates.
\end{methods}

\noindent
{\bf Data Availability}

All the data are publicly available through the ESO archive (\url{http://archive.eso.org/cms.html}). The data that supports the plots within this paper and other findings of this study
are available from the corresponding author upon reasonable request.

\begin{table}
\centering
\caption {Observing epochs for our SPHERE-IFS observations in 2015 and 2017, along with the SPHERE IRDIS (Avenhaus et al. 2018) and GPI (Rapson et al. 2015b) polarimetric datasets acquired in 2016 and 2014, respectively. The Barycentric Julian Date BJD (+2400000) and the central binary phase are given in Columns 2, 3, and 4 respectively. 
The shadow locations are shown in the last four columns for the inner and the outer rings (units are degrees), for the near and far side of the disk. Errors are 1-sigma.}
\vspace{0.8cm}
\begin{tabular}{lccccccr}
\hline\hline
Observing     & BJD  & Binary  & Binary  & Shadow$_{\rm near}$  & Shadow$_{\rm near}$  & Shadow$_{\rm far}$  & Shadow$_{\rm far}$  \\
  epoch         & (+ 2400000)           & phase & phase     &      ({\small inner ring})  &  ({\small outer ring}) & ({\small inner ring})  &  ({\small outer ring}) \\
                          &         &           &     (deg) & (deg) &  (deg) & (deg) & (deg) \\
\hline  
2014 & 56770.749 &  0.028      &  10.0   &        3.8$\pm$3.0  & ---- & 193.6$\pm$3.0 & ---- \\                         
2015 & 57146.831 &  0.350     &  126.1  &    122.9$\pm$0.6    &    117.3$\pm$0.6        & 284.5$\pm$1.1 & 276.6$\pm$1.1     \\
2016 & 57461.905 &  0.476     &  171.5  &    164.2$\pm$0.6     &     159.5$\pm$0.6      &  328.2$\pm$1.1 &  314.7$\pm$1.1 \\
2017 & 57870.879 &  0.384    &  138.1  &    134.6$\pm$0.6    &  128.3$\pm$0.6	&     305.6$\pm$1.1 & 291.3$\pm$1.1 	\\
\hline\hline
\end{tabular}
\end{table}

\clearpage

\begin{figure}
\includegraphics[width=1.00\columnwidth]{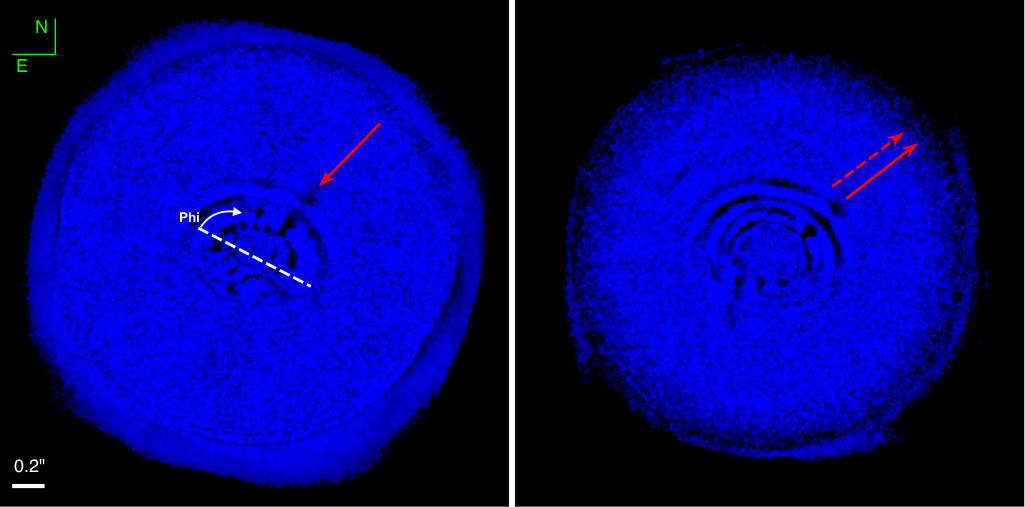}
\caption{ Comparison of IFS observations for 2015 (YJ bands) and 2017 (YH band) after a PCA algorithm (with 100 modes) has been applied to remove quasi-static speckles.  
The shadow location for the near side of the disk (indicated with a solid red arrow) has rotated (11$\pm$1 degrees), as expected from the binary orbital phase. The corresponding location of the shadow in the 2015 dataset is marked with a dashed arrow in the 2017 observations. The binary phase is defined as shown from the white line, setting $\phi$=0 from the semi-major axis of the disk. Note that $\phi$ has been measured on the de-projected image of the disk. The bright artefacts close to the shadow position are due to ADI/PCA post-processing technique but do not impact our analysis as we use the PDI images to measure the shadow depth (see Figure~3).}
\end{figure}

\begin{figure}
\includegraphics[width=1.00\columnwidth]{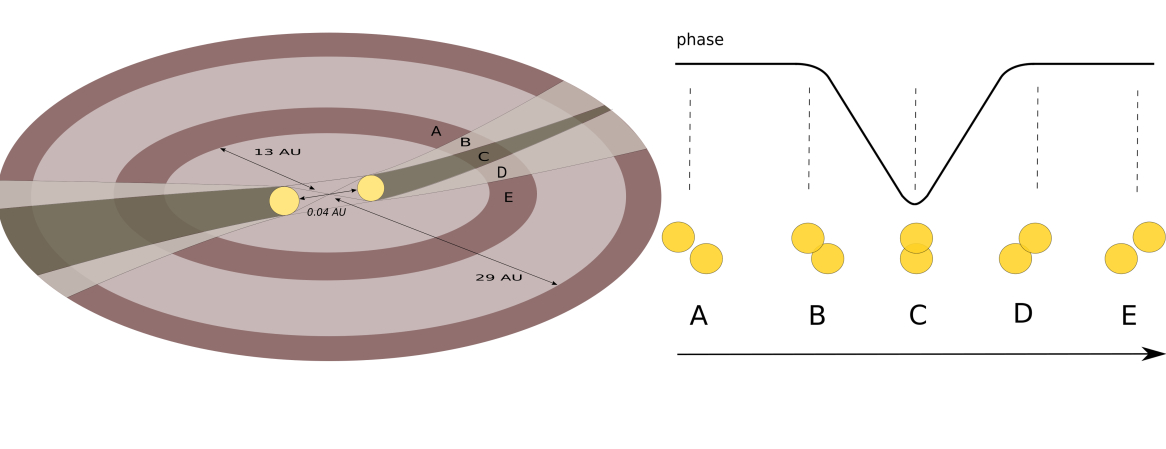}
\caption{Simplified sketch of the shadow phenomenon. Distances of the two stars and size of the disk are not in scale. 
Umbra (dark grey) and penumbra (light grey) due to the eclipses between the components 
are projected onto the disk with a distorted pattern due to the finite speed of light speed. The ring locations (13 and 29 au) and the binary separation are shown. 
The labelled phases of the eclipse phenomenon (A, B, C, D, E) corresponds to flux reduction as shown in the lower panel. }
\end{figure}

\begin{figure}
\includegraphics[width=0.44\columnwidth]{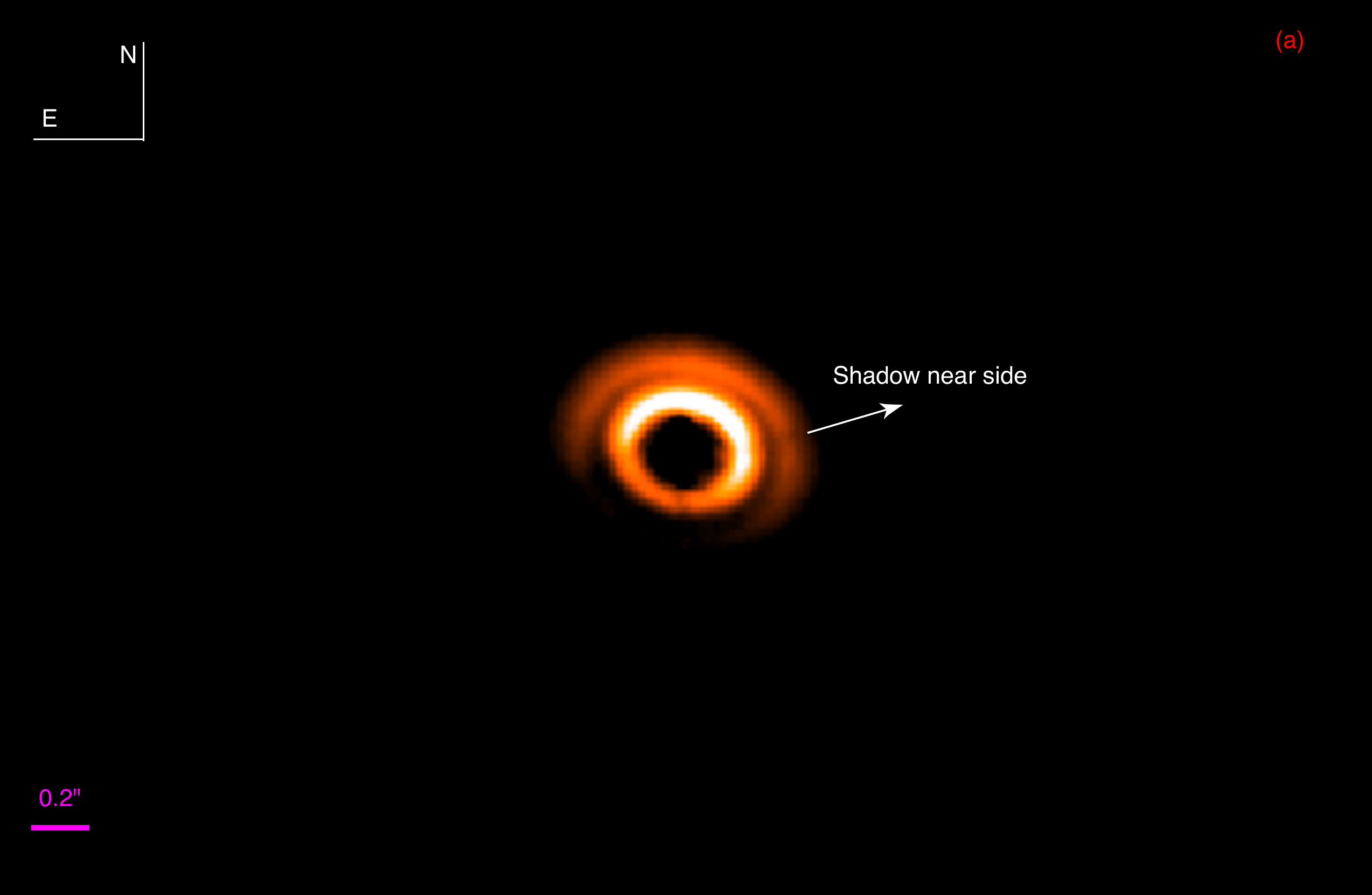}
\includegraphics[width=0.5\columnwidth]{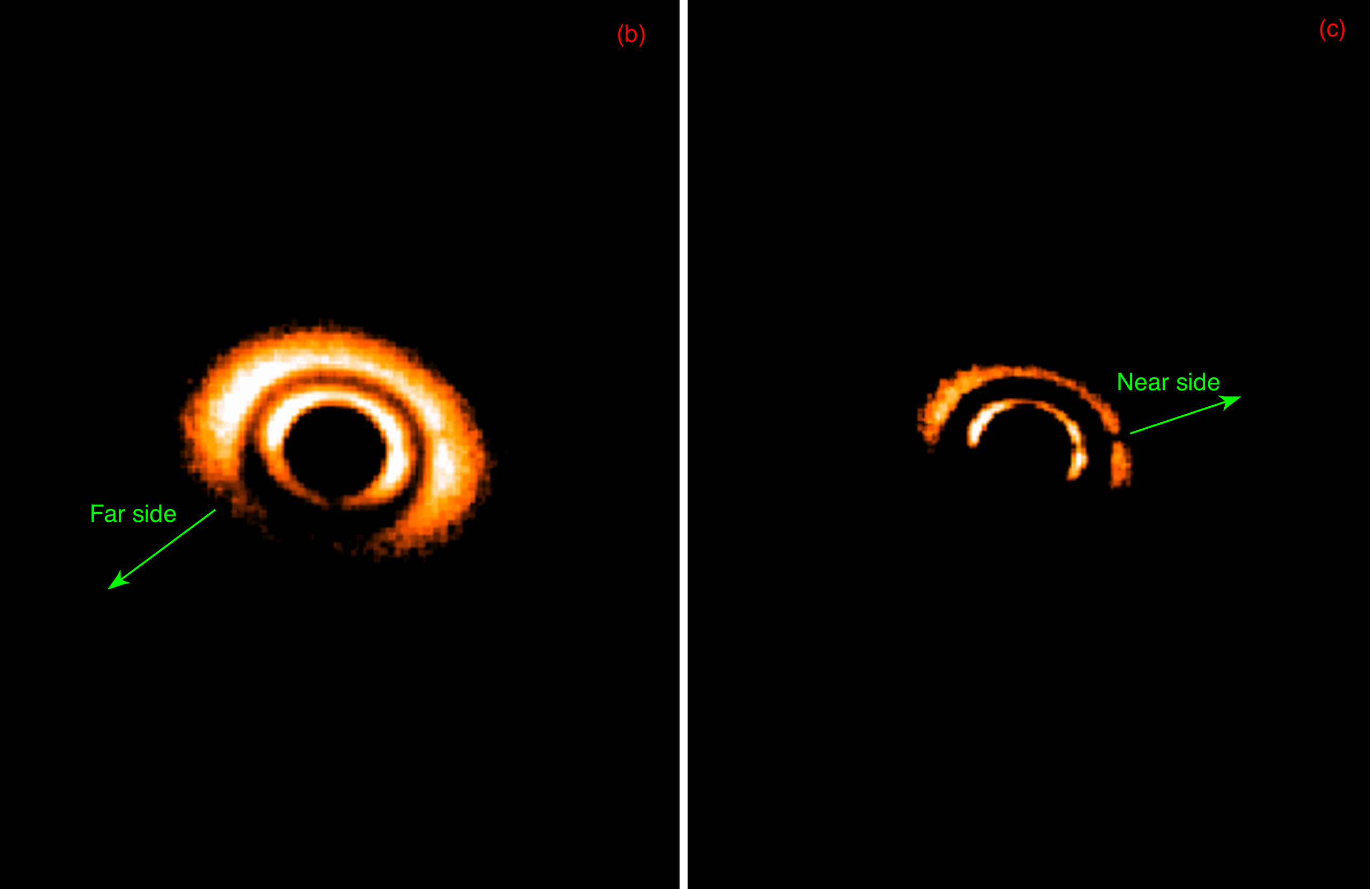}
\center
\includegraphics[width=0.4\columnwidth]{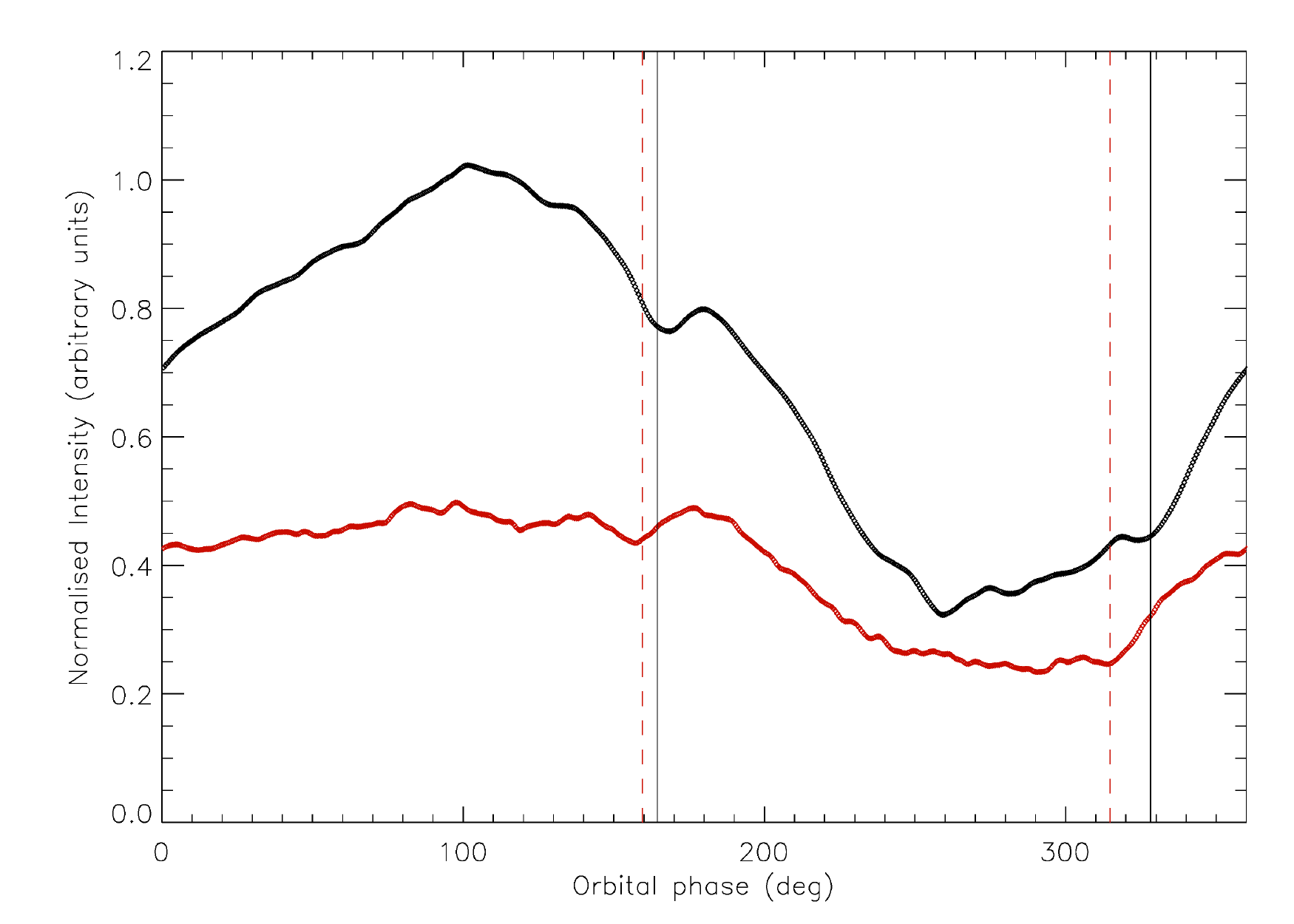}
\caption{Upper panel: Polarised IRDIS H band observations for V4046 Sgr \cite{avenhaus18} are shown in panel (a). An un-sharp masking algorithm has 
been applied to the original image to enhance the shadow locations, with different cuts for the near and far side 
(panels b and c, respectively). 
Each pixel image has also been multiplied by r$^2$ to compensate for the stellar illumination drop-off with radial distance.
Lower panel: Intensity profile of the disk $vs$ the orbital phase from PDI observations (IRDIS H band) for the circumbinary disk orbiting V4046 Sgr. 
The ring brightness variation along the position angle is due to scattering processes. The solid vertical lines indicate the expected positions of the shadows at the inner ring 
(black solid lines) for one shadow at  $\phi$=164.2$\pm$1.1 (near side of the disk) and the second one at $\phi$=328.2$\pm$1.1 deg 
(on the far side of the disk, see Method for $\phi$ definition). 
The error on the shadow position in the PDI dataset is calculated using a formula for Gaussian centering with dependency on 
FWHM, signal-to-noise ratio, and sampling. As expected, the difference in the position between the far side and near side of the disk is not exactly 
180 degrees due to disk flaring. The same holds for the outer ring (red dashed lines), where shadows are located at $\phi$=159.5$\pm$0.6 deg and $\phi$=314.7$\pm$1.1 deg.}
\end{figure}


\clearpage



\begin{thebibliography}{1}

\bibitem[1]{garufi} Garufi, A., Quanz, S.~P., Schmid, H.~M., Avenhaus, H., Buenzli, E., Wolf, S..\ Shadows and cavities in protoplanetary disks: HD 163296, HD 141569A, and HD 150193A in polarized light.\ {\it Astron. Astrophys.}, {\bf 568}, A40 (2014).

\bibitem[2]{marino15}  Marino, S., Casassus, S., Perez, S., Lyra, W., Roman, P. E., Avenhaus, H., Wright, C. M., Maddison, S. T. Compact Dust Concentration in the MWC 758 Protoplanetary Disk. {\it Astrophys. J.}, {\bf 813}, 76 (2015).

\bibitem[3]{stempels04} Stempels, H.~C., Gahm, G.~F.\ The close T Tauri binary V 4046 Sagittarii.\  {\it Astron. Astrophys.}, {\b 421}, 1159-1168 (2004).

\bibitem[4]{rosenfeld13}  Rosenfeld, K.~A., Andrews, S.~M., Wilner, D.~J., Kastner, J.~H., McClure, M.~K.\ The Structure of the Evolved Circumbinary Disk around V4046 Sgr.\ 
{\it Astrophys. J.}, {\bf 775}, 136 (2013).

\bibitem[5]{torres08} Torres, C.~A.~O., Quast, G.~R., Melo, C.~H.~F., Sterzik, M.~F.\ {\it Young Nearby Loose Associations.\ Handbook of Star Forming Regions}, {\bf II}, 5, 757 (2008).

\bibitem[6]{torres06} Torres, C.~A.~O., Quast, G.~R., da Silva, L., de La Reza, R., Melo, C.~H.~F., Sterzik, M.\ Search for associations containing young stars (SACY). I. Sample and searching method.\ {\it Astron. Astrophys.}, {\bf 460}, 695-708 (2006),

\bibitem[7]{wyatt08} Wyatt, M.~C.\ Evolution of Debris Disks.\ {\it Ann. Rev. Astron. Astrophys.}, {\bf 46}, 339 (2008).

\bibitem[8]{alexander12} Alexander, R.\ The Dispersal of Protoplanetary Disks around Binary Stars.\  {\it Astrophys. J.}, {\bf 757}, L29 (2012).

\bibitem[9]{merrill50} Merrill, P.~W., Burwell, C.~G.\ Additional Stars whose Spectra have a Bright H {$\alpha$} Line..\  {\it Astrophys. J.}, {\bf 112}, 72 (1950).

\bibitem[10]{jensen96} Jensen, E.~L.~N., Mathieu, R.~D., Fuller, G.~A.\ The Connection between Submillimeter Continuum Flux and Binary Separation in Young Binaries: Evidence of Interaction between Stars and Disks.\ {\it Astrophys. J.}, {\bf 458}, 312 (1996).

\bibitem[11]{macintosh14} Macintosh, B., and 46 colleagues.\ First light of the Gemini Planet Imager.\ {\it Proceedings of the National Academy of Science}, {\bf 111}, 12661 (2014).

\bibitem[12]{rapson15b} Rapson, V.~A., Kastner, J.~H., Andrews, S.~M., Hines, D.~C., Macintosh, B., Millar-Blanchaer, M., Tamura, M.\ Scattered Light from Dust in the Cavity of the V4046 Sgr Transition Disk.\  {\it Astrophys. J.}, {\bf 803}, L10 (2015).

\bibitem[13]{beuzit} Beuzit, J.-L., and 42 colleagues.\ SPHERE: a 'Planet Finder' instrument for the VLT.\ {\it Ground-based and Airborne Instrumentation for Astronomy}, {\bf II}, 7014 , 701418 (2008).

\bibitem[14]{claudi} Claudi, R.~U., and 14 colleagues.\ SPHERE IFS: the spectro differential imager of the VLT for exoplanets search.\ {\it Ground-based and Airborne Instrumentation for Astronomy}, {\bf II}, 7014, 70143E (2008).

\bibitem[15]{dohlen} Dohlen, K., and 15 colleagues.\ The infra-red dual imaging and spectrograph for SPHERE: design and performance.\ {\it Ground-based and Airborne Instrumentation for Astronomy}, {\bf II}, 7014, 70143L (2008).

\bibitem[16]{rosenfeld12} 	Rosenfeld, K.~A., Andrews, S.~M., Wilner, D.~J., Stempels, H.~C.\ A Disk-based Dynamical Mass Estimate for the Young Binary V4046 Sgr.\ {\it Astrophys. J.}, 
{\bf  759}, 119 (2012)

\bibitem[17]{benisty17} Benisty, M., and 48 colleagues.\ Shadows and spirals in the protoplanetary disk HD 100453.\ {\it Astron. Astrophys.}, {\bf 597}, A42 (2017)

\bibitem[18]{itoh14} Itoh, Y., and 49 colleagues.\ Near-infrared polarimetry of the GG Tauri A binary system.\ {\it Research in Astronomy and Astrophysics}, {\bf 14}, 1438-1446 (2014).


\bibitem[19]{wolff16} Wolff, S.~G., and 27 colleagues.\ The PDS 66 Circumstellar Disk as Seen in Polarized Light with the Gemini Planet Imager.\ {\it Astrophys. J.}, {\bf 818}, L15 (2016). 

\bibitem[20]{debes17} Debes, J.~H., and 10 colleagues.\ Chasing Shadows: Rotation of the Azimuthal Asymmetry in the TW Hya Disk.\  {\it Astrophys. J.}, {\bf 835}, 205 (2017).

\bibitem[21]{pohl15} Pohl, A., Pinilla, P., Benisty, M., Ataiee, S., Juh{\'a}sz, A., Dullemond, C.~P., Van Boekel, R., Henning, T.\ Scattered light images of spiral arms in marginally gravitationally unstable disks with an embedded planet.\ {\it Mon. Not. R. Astron. Soc.}, {\bf 453}, 1768-1778 (2015).

\bibitem[22]{min17} Min, M., Stolker, T., Dominik, C., Benisty, M.\ Connecting the shadows: probing inner disk geometries using shadows in transitional disks.\ 
{\it Astron. Astrophys.}, {\bf 604}, L10 (2017)

\bibitem[23]{facchini18} Facchini, S., Juh{\'a}sz, A., Lodato, G.\ Signatures of broken protoplanetary disks in scattered light and in sub-millimetre observations.\ 
{\it Mon. Not. R. Astron. Soc.}, {\bf 473}, 4459-4475 (2018).


\bibitem[24] {avenhaus18} Avenhaus, H., Quanz, S.~P., Garufi, A., Perez, S., Casassus, S., Pinte, C., Bertrang, G.~H.-M., Caceres, C., Benisty, M., Dominik, C.\ Disks around T Tauri Stars with SPHERE (DARTTS-S). I. SPHERE/IRDIS Polarimetric Imaging of Eight Prominent T Tauri Disks.\   {\it Astrophys. J.}, {\bf 863}, 44 (2018).

\bibitem[25] {kuruwita18} Kuruwita, R.~L., Ireland, M., Rizzuto, A., Bento, J., Federrath, C.\ Multiplicity of disk-bearing stars in Upper Scorpius and Upper Centaurus-Lupus.\ 
{\it Mon. Not. R. Astron. Soc}, {\bf 480}, 5099-5112 (2018).

\bibitem[26] {chauvin17} Chauvin, G., Desidera, S., Lagrange, A.-M., Vigan, A., Feldt, M., Gratton, R., Langlois, M., Cheetham, A., Bonnefoy, M., Meyer, M.\ SHINE, The SpHere INfrared survey for Exoplanets.\ {\it SF2A-2017: Proceedings of the Annual meeting of the French Society of Astronomy and Astrophysics}, 331-335 (2017).

\bibitem[27]{vigan10} Vigan, A., Moutou, C., Langlois, M., Allard, F., Boccaletti, A., Carbillet, M., Mouillet, D., Smith, I.\ Photometric characterization of exoplanets using angular and spectral differential imaging. {\it Mon. Not. R. Astron. Soc.} {\bf 407}, 71-82 (2010).

\bibitem[28] {pavlov} Pavlov, A., M{\"o}ller-Nilsson, O., Feldt, M., Henning, T., Beuzit, J.-L., Mouillet, D.\ SPHERE data reduction and handling system: overview, project status, and development.\ {\it Advanced Software and Control for Astronomy II}, {\bf 7019}, 701939 (2008).

\bibitem[29] {delorme17} Delorme, P., and 19 colleagues.\ The SPHERE Data Center: a reference for high contrast imaging processing.\ {\it SF2A-2017: Proceedings of the Annual meeting of the French Society of Astronomy and Astrophysics}, 347-361 (2017). 

\bibitem[30]{maire16} Maire, A.-L., and 18 colleagues.\ SPHERE IRDIS and IFS astrometric strategy and calibration.\ {\it Ground-based and Airborne Instrumentation for Astronomy VI}, {\bf 9908}, 990834 (2016).

\bibitem [31] {marois06} Marois, C., Lafreni{\`e}re, D., Doyon, R., Macintosh, B., Nadeau, D.\ Angular Differential Imaging: A Powerful High-Contrast Imaging Technique.\ 
{\it Astrophys. J.}, {\bf 641}, 556-564 (2006).

\bibitem[32]{galicher11} Galicher, R., Marois, C., Macintosh, B., Barman, T., Konopacky, Q.\ M-band Imaging of the HR 8799 Planetary System Using an Innovative LOCI-based Background Subtraction Technique.\ {\it Astrophys. J.}, {\bf 739}, L41 (2011).

\bibitem[33] {mesa15} Mesa, D., and 35 colleagues.\ Performance of the VLT Planet Finder SPHERE. II. Data analysis and results for IFS in laboratory.\ 
{\it Astron. Astrophys.}, {\bf 576}, A121 (2015).

\bibitem[34] {galicher18} Galicher, R., and 23 colleagues. \ Astrometric and photometric accuracies in high contrast imaging: The SPHERE speckle calibration tool (SpeCal).\ 
{\it Astron. Astrophys.}, {\bf 615}, A92 (2018).

\bibitem[35] {avenhaus14} Avenhaus, H., Quanz, S.~P., Schmid, H.~M., Meyer, M.~R., Garufi, A., Wolf, S., Dominik, C.\ Structures in the Protoplanetary Disk of HD142527 Seen in Polarized Scattered Light.\  {\it Astrophys. J.}, {\bf 781}, 87 (2014).

\bibitem[36]{quast00} Quast, G.~R., Torres, C.~A.~O., de La Reza, R., da Silva, L., Mayor, M.\ V4046 Sgr, a key young binary system..\ {\it IAU Symposium 200}, {\bf 28} (2000).

\bibitem[37] {donati11} Donati, J.-F., and 14 colleagues.\ The close classical T Tauri binary V4046 Sgr: complex magnetic fields and distributed mass accretion.\ 
{\it Mon. Not. R. Astron. Soc}, {\bf 417}, 1747-1759 (2011). 

\bibitem [38] {milli12} Milli, J., Mouillet, D., Lagrange, A.-M., Boccaletti, A., Mawet, D., Chauvin, G., Bonnefoy, M.\ Impact of angular differential imaging on circumstellar disk images.\ 
{\it Astron. Astrophys.}, {\bf 545}, A111 (2012).

\bibitem [39] {stolker16} Stolker, T., Dominik, C., Min, M., Garufi, A., Mulders, G.~D., Avenhaus, H.\ Scattered light mapping of protoplanetary disks.\ 
{\it Astron. Astrophys.}, {\bf 596}, A70 (2016). 

\bibitem [40] {sissa18} Sissa, E., and 64 colleagues. 
\ High-Contrast study of the candidate planets and protoplanetary disk around HD\~{}100546.\ ArXiv e-prints arXiv:1809.01001 (2018).

\bibitem[41]{maire17} Maire, A.-L., and 44 colleagues.\ Testing giant planet formation in the transitional disk of SAO 206462 using deep VLT/SPHERE imaging.\ 
{\it Astron. Astrophys.}, {\bf 601}, A134 (2017).

\bibitem [42] {henyey} Henyey, L.~G., Greenstein, J.~L.\ Diffuse radiation in the Galaxy.\ {\it Astrophys. J.} {\bf 93}, 70-83 (1941).

\bibitem [43] {milli17} Milli, J., and 30 colleagues.\ Near-infrared scattered light properties of the HR 4796 A dust ring. A measured scattering phase function from 13.6 deg to 166.6 deg.\ 
{\it Astron. Astrophys.}, {\bf 599}, A108 (2017).

\bibitem[44]{kama16} Kama, M., Pinilla, P., Heays, A.~N.\ Spirals in protoplanetary disks from photon travel time.\ {\it Astron. Astrophys.}, {\bf 593}, L20 (2016). 


\end{thebibliography}
\end{document}